\documentclass[conference]{IEEEtran}

\usepackage[T1]{fontenc}
\usepackage{cite}
\usepackage{graphicx}
\usepackage{epstopdf}
\usepackage{amssymb}
\usepackage{amsmath}
\usepackage{bbm}
\usepackage{multirow}
\usepackage{multicol}
\usepackage{subfigure}
\usepackage[ruled,vlined]{algorithm2e}
\usepackage{bbm}
\usepackage{array}
\usepackage{empheq}
\usepackage{dblfloatfix}
\usepackage{placeins}
\usepackage{booktabs}
\usepackage{color}
\usepackage{amsthm}
\begin{document}
\title{User Detection Performance Analysis for \\Grant-Free Uplink Transmission in \\Large-Scale Antenna Systems}

\author{
	\IEEEauthorblockN{Jonghyun Kim, Kyung Lin Ryu, and Kwang Soon Kim}%
	\IEEEauthorblockA{School of Electrical and Electronic Engineering, Yonsei University\\ 50 Yonsei-ro, Seodaemun-gu, Seoul, 03722, Korea.\\Email: jonghyun.kim@yonsei.ac.kr, kyunglin.ryu@yonsei.ac.kr, ks.kim@yonsei.ac.kr}
}
\markboth{Journal of \LaTeX\ Class Files,~Vol.~PP, No.~99, October~2017}%
{Shell \MakeLowercase{\textit{et al.}}: Bare Demo of IEEEtran.cls for IEEE Communications Society Journals}

\maketitle

\begin{abstract}
In this paper, user detection performance of a grant-free uplink transmission in a large scale antenna system is analyzed, in which a general grant-free multiple access is considered as the system model and Zadoff-Chu sequence is used for the uplink pilot. The false alarm probabilities of various user detection schemes under the target detection probabilities are evaluated.
\end{abstract}

\begin{IEEEkeywords}
Grant-free uplink transmission, large-scale antenna system, Neyman-Pearson decision criterion, ultra-reliable low-latency communication.
\end{IEEEkeywords}

\section{Introduction} 

To enable the diverse future wireless services, the 5th generation (5G) mobile communication has been extensively discussed, in which the 5G services are categorized as enhanced mobile broadband (eMBB), massive machine type communications (mMTC), and ultra-reliable low-latency communications (URLLC) according to the target requirements for the services.
Among them, the URLLC can support so called \emph{mission-critical control} services, including automated driving, tele-surgery (e-health), video-driven interaction, augmented reality applications, and factory automations \cite{5GUC}, which typically requires high reliability level of 99.999\% and 1 ms end-to-end latency.
Such services require various kinds of packets with different traffic characteristics, including packet size, arrival rate and model, etc.
In this paper, we focus on \emph{sporadic event-driven uplink transmission} in which traffics occur randomly but require high reliability, low latency, and medium to high throughput, which would be one of the most challenging tasks.

In the previous 4G communication (LTE-Advanced), the uplink random access is \emph{grant-based} so that a 4-way handshaking must be completed before a data transmission from user equipments (UEs) to a base station (BS).
The 4-way handshaking includes 1) random access preambles from UEs, 2) random access responses from a BS, 3) scheduling requests or buffer status reports from UEs, and then 4) uplink scheduling grants from a BS \cite{3GPP.36.321}.
Through theses 4 steps, a BS recognizes the requests for uplink transmissions from UEs, identifies which UEs will transmit, and allocates resources to them to establish uplink connections.
But this procedures cause severe delays at least 4 times of the transmission time interval (TTI) and decrease the spectral efficiency, especially as the latency requirement becomes tight.
Also, the hybrid automatic repeat request (H-ARQ) employed for achieving reliability cannot work appropriately due to the low latency requirement.
To overcome these limitations, a \emph{grant-free} multiple access (GFMA) has been proposed and discussed \cite{R1-1608757}.
For such a GFMA, one important task is to design a highly reliable user detection scheme \cite{R1-1608855} because a BS does not have any prior information about which UEs will transmit and also its performance should be reliable enough to meet the high reliability requirements of a URLLC.

Recently, a large-scale antenna system (LSAS), which employs massive antennas at a BS, has been in a great interest due to its inherit merit of providing massive spatial dimension \cite{TWC.2010.Marzetta}.
To enhance the user detection performance in a GFMA, a large spatial multiplexing gain as well as a large diversity gain can be utilized.
Moreover, as the number of antennas increases, due to the \emph{channel hardening effect} \cite{TIT.2004.Hochwald}, we can predict the achievable data rate without considering the instantaneous channel state information and therefore it is possible to schedule users by using only their long-term channel states \cite{ICTC.2016.Kim,ACC.2017.Choi}.

\section{System Model}

\begin{figure}
	\centering
	\includegraphics[width=7.5cm]{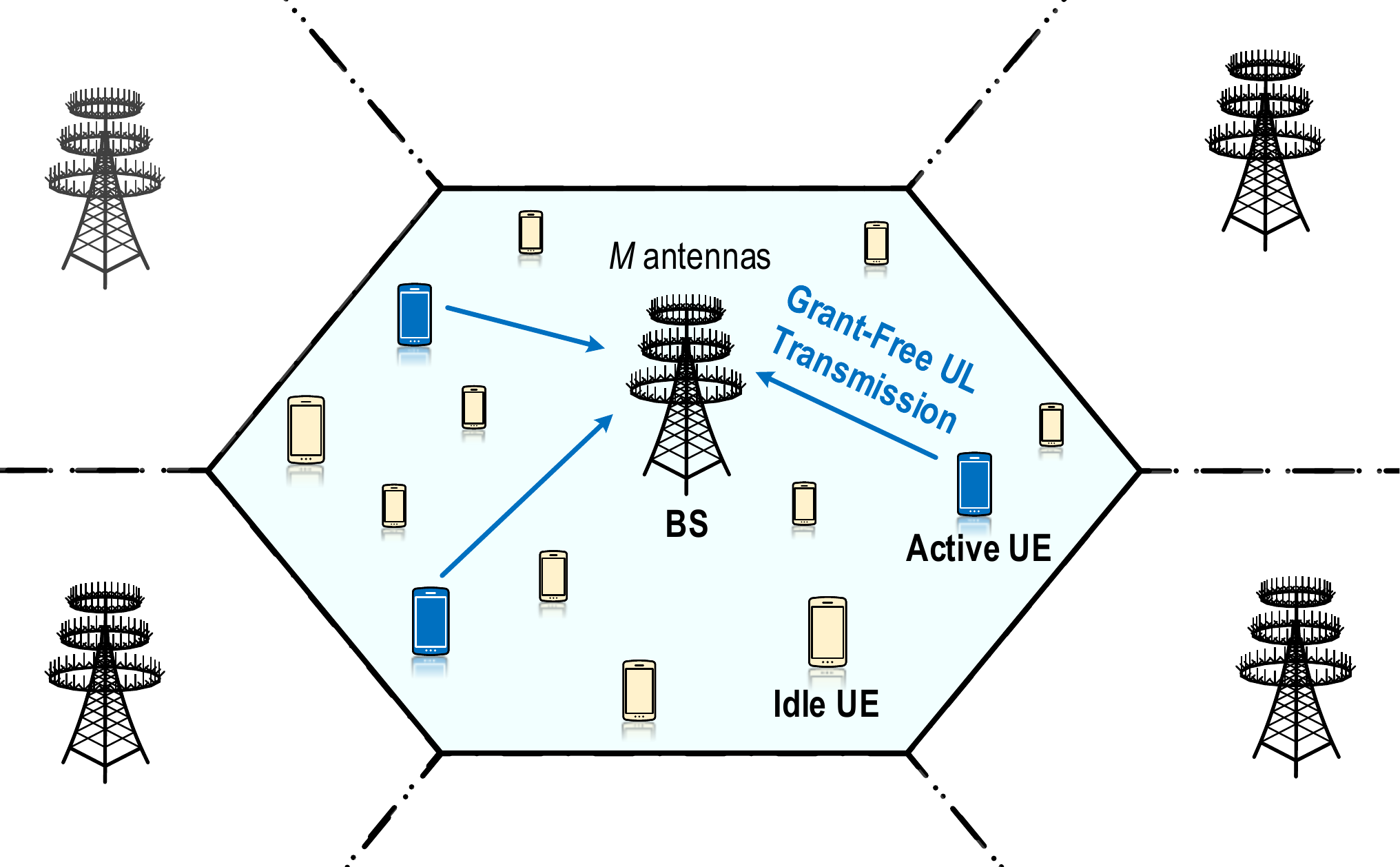}
	\caption{A grant-free multiple access system model}
	\label{fig.sysmodel}
\end{figure}

\subsection{Grant-Free Multiple Access}

We consider an uplink large-scale antenna system as illustrated in Fig. \ref{fig.sysmodel}, where each of BS is equipped with $M$ antennas and each of $U$ UEs is equipped with a single antenna.
Also, we configure the time-frequency sub-channels with a sub-frame with a duration of $T_s$ and the sub-band bandwidth $W_s$.
Within each sub-channel, the channel state is assumed to be static so that the $M\times 1$ SIMO spatial small scale block fading, denoted by ${\bf h}_j$, does not change, while the large scale fading ${\beta}_j$ only depends on the location of UE and does not change also within the scheduling period.
We assume a perfect non line of sight (NLOS) propagation so that the small scale fading is modeled by a Rayleigh distribution.
The UEs are divided into scheduling groups and then the BS allocates the sub-channels to the scheduling groups.
The UEs allocated in the same sub-channel access to the BS by using space division multiple access (SDMA) and every UE with a new packet arrival gets activated and transmits by using the allocated sub-channel with a size of $N=W_sT_s$ symbols.
For analytical tractability, it is assumed that the packet arrival follows a Poisson distribution with the average arrival rate of $P_A$.
The BS and UE are assumed to operate in a perfect synchronization.

\subsection{Channel and Signal Model}

Let ${\cal O}$ be the set of users scheduled at the sub-channel of the sub-frame $t$ and the sub-band $f$.
Denote ${\bf x}_j$ the transmission block of user $j$ in ${\cal O}$, then the $M \times N$ received signal matrix ${\bf Y}$ at the BS of interest can be expressed as
\begin{align}
{\bf Y}= \sum\limits_{j\in {\cal O}} \sqrt{\beta_j} {\bf h}_j{\bf x}_j^{\rm H} {\mathbbm 1}^{}_j + {\bf V},
\end{align}
where ${\mathbbm 1}_{j}$ is the indicator function for the activation of user $j$ and ${\bf V}$ is the $M \times N$ matrix for the thermal noise and other-cell interference signals.
The transmission block ${\bf x}_{j}$ has $L$ symbols for its pilot sequence and $N-L$ symbols for data so that we can write ${\bf x}_{j}$ as $ {\bf x}_{j}
= \left[\sqrt{p_j^{\rm tr}} \, {\boldsymbol \psi}_{j}^{\rm H},\:\sqrt{p_j^{\rm dt}} \, {\bf d}_j^{\rm H}\right]^{\rm H}$,
where ${\boldsymbol \psi}_{j}$ denotes the pilot sequence of length $L$ and ${\bf d}_j$ denotes the $N-L$ data symbol vector.
Also, $p_j^{\rm tr}$ and $p_j^{\rm dt}$ denote the powers of the pilot sequence and the data symbols, respectively.
Here we assume that $p_j^{\rm tr}$ is determined by the full channel inversion power control for a given common target received power $\overline{p}^{\rm tr}$, i.e., $p_j^{\rm tr} = \overline{p}^{\rm tr}\beta_j^{-1}$.
Let ${\bf Y} = \left[{\bf Y}^{\rm tr} , {\bf Y}^{\rm dt} \right]$ and ${\bf V} = \left[{\bf V}^{\rm tr} , {\bf V}^{\rm dt} \right]$, then ${\bf Y}^{\rm tr}$ can be written
\begin{align}
	{\bf Y}^{\rm tr} = \sum\limits_{j\in {\cal O}} \sqrt{\overline{p}^{\rm tr}} {\bf h}_j{\boldsymbol \psi}_j^{\rm H} {\mathbbm 1}^{}_j + {\bf V}^{\rm tr}.
\end{align}
For the pilot sequence, we use Zadoff-Chu (ZC) sequences currently used in LTE-A \cite{3GPP.36.211}.   
With a length of a prime number $L$, there are sequences with $L-1$ different roots and $L$ circularly shifted versions for each root so that there are $L^2-L$ sequences available.
Note that the cross correlation value is zero between the ZC sequences of circularly shifted versions in the same root and is a constant of $\tfrac{1}{L}$ between sequences from different roots.
As the cross correlation causes interferences and degrades the user detection performance, the sequence assignment must maximize the number of sequences of the same root.
Therefore, when $K=|{\cal O}|$ and the sequence length is $L$, the number of used roots is minimized to $\left\lceil\tfrac{K}{L}\right\rceil$.
For example, the pilot symbol for user $j$ is assigned as ${\psi}_{j}(l) = {\left< \exp\left(- {\sf j}\pi {{{r_j}l(l+1)}}/{L}\right) \right>}_{s_j}$,
where the root is given as $r_j = 1 + \mod \left( j-1 ,\left\lceil\tfrac{K}{L}\right\rceil \right)$, the circularly shift is given as $s_j = 1 + \left\lfloor(j-1)/\left\lceil\tfrac{K}{L}\right\rceil\right\rfloor$, and $\left<\cdot\right>_s$ denotes the circular shift by $s$ symbols.
Let define ${\cal O}_{r_j}$ be the set of users in ${\cal O}$ whose pilot sequences have root $r_j$ and let $K_{r_j} = |{\cal O}_{r_j}|$.

\section{The Neyman-Pearson Decision Criterion}

The number of activated users in ${\cal O}$ at a time follows a binomial distribution ${\cal B}(K,P_A)$ because we assume a Poisson random packet arrival.
At the BS side, for every sub-frame duration, it tries to detect which users transmitted signals by using the sequences and a correlation detector.
Let define $Z_j = \left\| {\bf Y}^{\rm tr} {\boldsymbol \psi}_j\right\|^2$, then $Z_j$ is the sufficient statistic for detecting user $j$ from the received signal ${\bf Y}$ by using sequence ${\boldsymbol \psi}_j$.
Also, define ${\widetilde{\bf v}}_j^{\rm tr} = {\bf V}^{\rm tr}{\boldsymbol \psi}_j $.
Then the sufficient statistic $Z_j$ can be written as
\begin{align} 
\begin{split} \label{eq.Z}
Z_j
= L\overline{p}^{\rm tr} \left\|{\bf h}_j  {\mathbbm 1}_j  \right\|^2 
+  \sum\limits_{k \in {\cal O} \backslash {\cal O}_{r_j}} \overline{p}^{\rm tr}\left\| {\bf h}_k  {\mathbbm 1}_k \right\|^2 
+ \left\| {\widetilde{\bf v}}_j^{\rm tr}\right\|^2 ,
\end{split}
\end{align} 
where ${\boldsymbol \psi}_j^{\rm H} {\boldsymbol \psi}_j  = L $, $\|{\boldsymbol \psi}_j^{\rm H} {\boldsymbol \psi}_k \|= 1$ if $r_j \neq r_k$, and $\|{\boldsymbol \psi}_j^{\rm H} {\boldsymbol \psi}_k \|= 0$ if $r_j = r_k,~j \neq k$.
For the second term in (\ref{eq.Z}), let define $J=\sum_{k \in {\cal O} \backslash {\cal O}_{r_j}}{\mathbbm 1}_k$, then $J \sim {\cal B} (K-K_{r_j},P_A)$.
Because ${\bf h}_j,{\bf h}_k \sim {\mathcal{CN}} ({\bf 0},{\bf I}_M)$ and ${\widetilde{\bf v}}_j^{\rm tr} \sim {\mathcal{CN}}({\bf 0},L^2{\bf I}_M)$, $Z_j$ follows a chi-square distribution with the degree of freedom (DoF) of $2M$ if $J$ is determined.
More specifically, $Z_j \sim \sigma^2_{J,{\rm active}}\chi_{2M}^2$ if ${\mathbbm 1}_j = 1$ and $Z_j \sim \sigma^2_{J,{\rm idle}}\chi_{2M}^2$ if ${\mathbbm 1}_j = 0$, where $\sigma^2_{J,{\rm active}} = (L\overline{p}^{\rm tr}+J\overline{p}^{\rm tr}+L^2)/2$ and $\sigma^2_{J,{\rm idle}} = (J\overline{p}^{\rm tr}+L^2)/2$.
In fact, due to the large value of $M$, $Z_j$ can be treated as a Gaussian random variable by using the central limit theorem.

By the NP decision criterion under a target detection probability $P_D$, the detection threshold $\Omega$ is the minimum value satisfying the detection probability condition as
\begin{align} 
\begin{split} \label{eq.P_D}
1-P_D &\ge P(Z_j\le\Omega|{\mathbbm 1}_j = 1) \\
&= \sum\limits_{q=0}^{K-K_{r_j}} P\left(Z_j\le\Omega|{\mathbbm 1}_j=1,J=q\right)P\left(J=q\right) \\
&= \sum\limits_{q=0}^{K-K_{r_j}} Q\left( \tfrac{\Omega/\sigma^2_{q,{\rm active}}-2M}{\sqrt{4M}} \right) P\left(J=q\right) ,
\end{split}
\end{align}
where $Q(\cdot)$ denotes the tail probability of the standard normal distribution.
Note that the decision criterion with a threshold of $\Omega$ in (\ref{eq.P_D}) can cause false alarm and the false alarm probability $P_{FA}$ can be calculated as
\begin{align}
\begin{split}
	P_{FA} &= P(Z_j>\Omega|{\mathbbm 1}_j = 0) \\
	&= \sum\limits_{q=0}^{K-K_{r_j}} P\left(Z_j>\Omega|{\mathbbm 1}_j=0,J=q\right)P\left(J=q\right) \\
	&= 1-\sum\limits_{q=0}^{K-K_{r_j}} Q\left( \tfrac{\Omega/\sigma^2_{q,{\rm idle}}-2M}{\sqrt{4M}} \right) P\left(J=q\right) .
\end{split}
\end{align}
Note that there always exists a threshold that satisfies the detection probability condition.
However, it is generally impossible to satisfy both the detection probability condition and the false alarm probability condition simultaneously.
The resulted false alarm probability by the threshold from the detection probability condition shows the distance between $P(Z_j=z|{\mathbbm 1}_j = 1)$ and $P(Z_j=z|{\mathbbm 1}_j = 0)$ and thus it can evaluate a trade-off between the detection probability and the false alarm probability.

\section{Numerical Evaluations}

The false alarm probabilities of various detection schemes under various target detection probabilities are evaluated numerically in this section.
For one sub-channel, the sub-frame duration is 1 ms and the sub-band bandwidth is 125 KHz.
Using orthogonal frequency division multiplexing (OFDM) waveform with 25\% cyclic prefix (CP) overhead, $N = 100$ symbols are used in a transmission block.
The target received power $\overline{p}^{\rm tr}$ is set to be 15 dB over the thermal noise and other-cell interference.
Here, we consider the maximum scheduling size constrained by the packet arrival rate $P_A$, the BS antenna size $M$, and the outage probability $P_O$.
Because, when ZF receiver is used at the receiver, the maximum supportable number of users is at most $M-1$ to have non-zero rates \cite{TIT.2004.Hochwald}.
Thus, the maximum scheduling size is $K_{\rm max} = \max~K,$ subject to $P(J\le M-1) \le P_O$, where $J \sim {\cal B}(K,P_A)$.
In addition, we used sufficiently large pilot length $L$ so that $L^2-L \ge K_{\rm max}$ for the simulations.

\begin{figure}
	\centering
	\includegraphics[width=7.3cm]{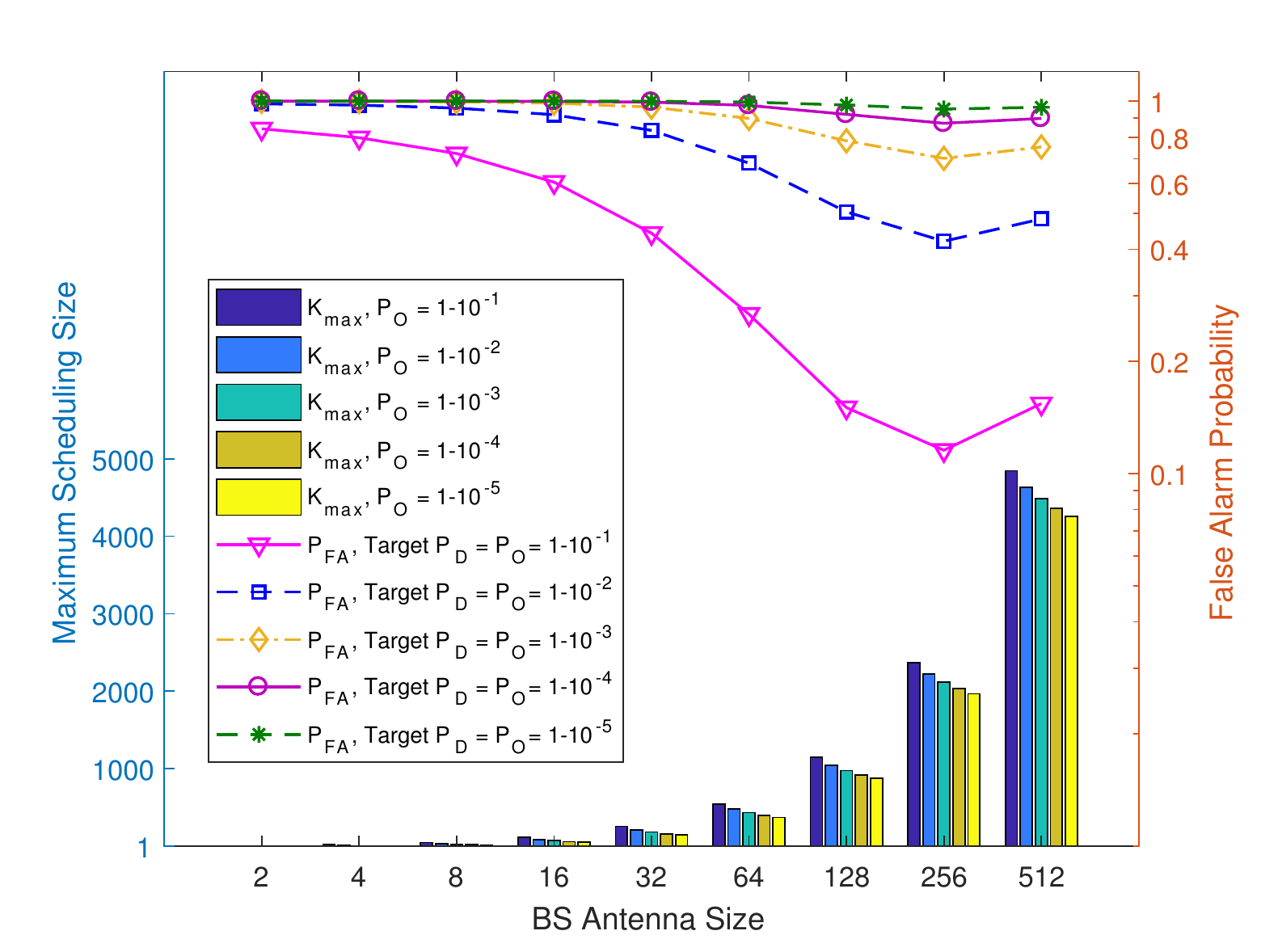}
	\caption{The false alarm probability and the maximum scheduling size versus the BS antenna size when $L=97$ and $P_A=0.1$}
	\label{antenna-vs-fa}
\end{figure}
\begin{figure}
	\centering
	\includegraphics[width=7.3cm]{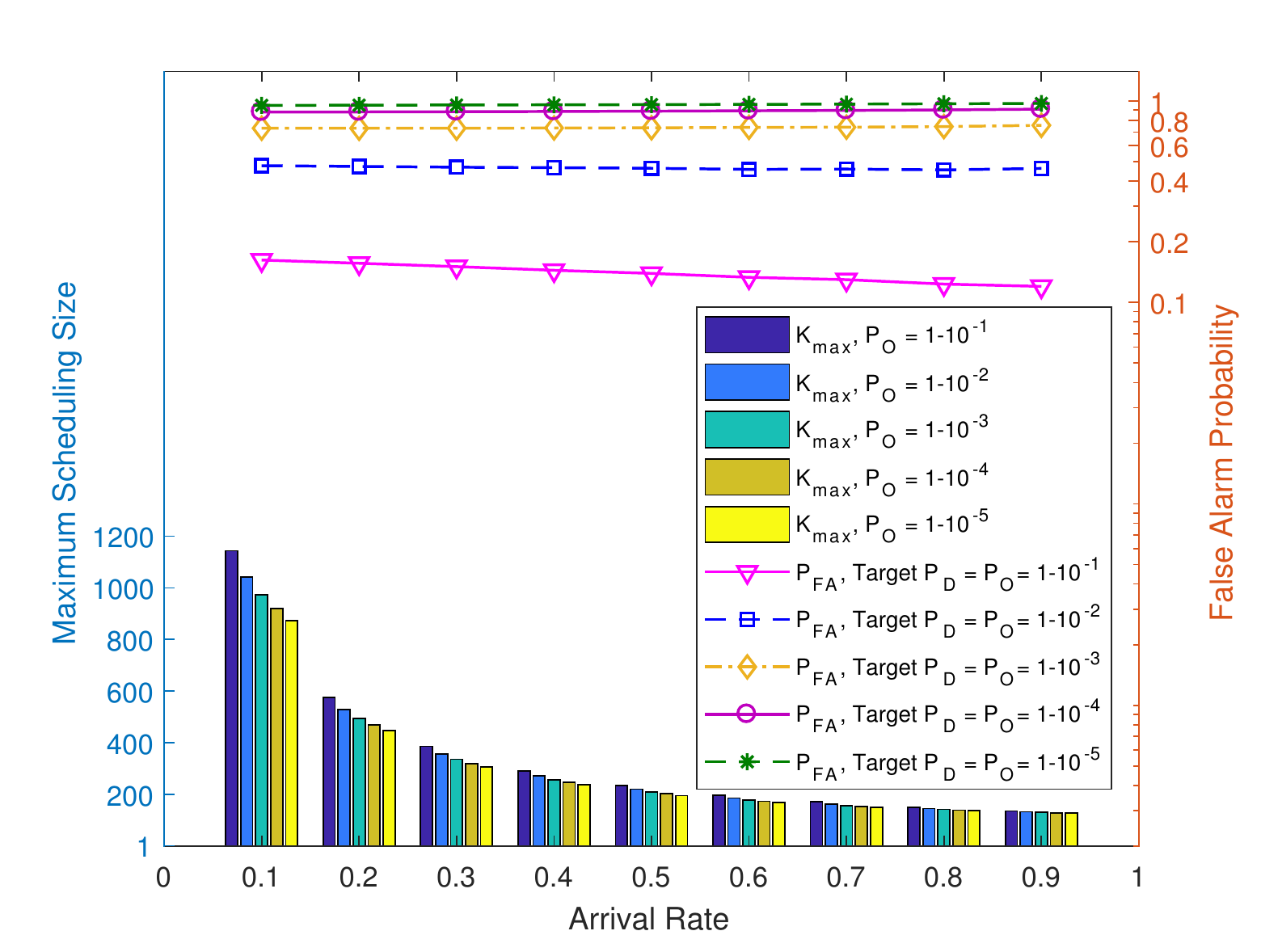}
	\caption{The false alarm probability and the maximum scheduling size versus the arrival rate when $M=128$ and $L=47$}
	\label{pa-vs-fa}
\end{figure}	

In Fig. \ref{antenna-vs-fa}, the false alarm probability and the maximum scheduling size are plotted.
Here, we set the pilot length $L=97$ because $K_{\rm max}$ increases up to 4846 as $M$ increases to 512.
In this figure, the false alarm probability dose not always decrease with the number of BS antennas.
The reason is that as the number of antennas increases, the number of interferers also increases very rapidly.
From this, the number of BS antennas at a given target detection probability can be seen to have some optimal value for a given false alarm probability.
In Fig. \ref{pa-vs-fa}, the false alarm probability and the maximum scheduling size are shown when the arrival rate $P_A$ changes from $0.1$ to $0.9$.
Here, since the maximum scheduling size decreases significantly while the arrival rate increases, the number of active users at a time does not change a lot, and thus the false alarm probability decreases very slightly.

\section{Conclusion}

In this paper, user detection schemes are analyzed in a general GFMA system model.
Assuming a Poisson packet arrival and using ZC sequences, the detection threshold for the target detection probability and the resulted false alarm probability are derived.
By employing a number of antennas at a BS, the resulted false alarm probability can be lowered generally to an acceptable level.
These results show that highly reliable detection schemes in GFMA for the future URLLC services are feasible in realistic LSAS environments.

\section*{Acknowledgment}
This work was supported by Institute for Information \& communications Technology Promotion (IITP) grant funded by the Korea government (MSIT) (2015-0-00300, Multiple Access Technique with Ultra-Low Latency and High Efficiency for Tactile Internet Services in IoT Environments) and (2014-0-00552, Next Generation WLAN System with High Efficient Performance).

\bibliographystyle{IEEEtran}
\bibliography{IEEEabrv,bibICUFN18}

\end{document}